\documentclass[pra,twocolumn,superscriptaddress]{revtex4}
\usepackage{color}

\usepackage{graphicx}
\begin{document}

\title{The low-energy excitation spectrum of one-dimensional dipolar quantum gases}

\author{S. De Palo}
\affiliation{DEMOCRITOS INFM-CNR and Dipartimento di Fisica Teorica,
  Universit\`a di  Trieste, Trieste, Italy }
\author{E. Orignac}
\affiliation{Universit\'e de Lyon, Lyon, France}
\affiliation{Laboratoire de Physique de l'\'Ecole Normale
  Sup\'erieure de Lyon, CNRS-UMR5672, Lyon, France}
\author{R. Citro}
\affiliation{Dipartimento di Fisica "E. R. Caianiello" and CNISM, Universit\`a degli Studi di Salerno, Salerno, Italy}
\affiliation{LPM$^2$C, CNRS, Grenoble, France}
\author{M.~L. Chiofalo}
\affiliation{INFN, Dpt. of Mathematics and Faculty of Pharmacy, University of Pisa, Pisa, Italy}
\affiliation{Classe di Scienze, Scuola Normale Superiore, Pisa, Italy}

\begin{abstract}
We determine the excitation spectrum of a bosonic dipolar quantum gas
in a one-dimensional geometry, from the dynamical density-density
correlation functions simulated by means of Reptation Quantum Monte
Carlo techniques. The excitation energy is always vanishing at the 
first vector of the reciprocal lattice in the whole 
crossover from the liquid-like at low density to the quasi-ordered
state at high density, demonstrating the absence of a roton minimum. 
Gaps at higher reciprocal lattice vectors are
seen to progressively close with increasing density, while the quantum
state evolves into a quasi-periodic structure. The simulational data
together with the uncertainty-principle inequality also provide a
rigorous proof of the absence of long-range order in such a super-strongly
correlated system. Our conclusions confirm that the dipolar gas is in
a Luttinger-liquid state, significantly affected by the dynamical
correlations. The connection with ongoing experiments is also discussed.

\end{abstract}
\pacs{ 03.75.Kk, 03.75.Hh, 71.10.Pm, 02.70.Ss, 31.15.Ew}
\maketitle

{\it Introduction.-}
Ultracold quantum gases with dipolar interactions are currently being produced in
laboratory, where atomic $^{52}$Cr atoms have been
Bose-condensed~\cite{cr_dipolar,cr_bec}, following earlier theoretical
predictions~\cite{dipolar_theo_earlier_1,dipolar_theo_earlier_2}.
Experiments have been suggested~\cite{zoller_buchler} aimed to produce
molecular gases with large dipolar strengths, and a few laboratories
worldwide are working along these lines. In fact,  
dipolar quantum gases are emerging as competitive realizations of
quantum devices~\cite{rabl_zoller} and as a
laboratory for investigating strongly correlated
regimes~\cite{dipolar_citro_pra_rc,astra} and novel quantum
phases~\cite{applications_1,applications_2}, 
in which quantum fluctuations are enhanced by exploiting 
techniques acquired for an accurate manipulation of atomic
gases. These include the possibility of lowering the temperature; of
tuning the interactions in both their long-range
tail~\cite{giovanazzi_pfau}
and in the strength of their short-range
part~\cite{pfau_a0} by means of the Fano-Feshbach
mechanism~\cite{ff_1,ff_2} to let emerge the dipolar character~\cite{dipolar_theo_recent_2}; 
of reducing the dimensionality down to one (1D), as already performed in other
systems~\cite{Bloch,Weiss}.  

One-dimensional quantum gases are naturally inclined to be strongly 
correlated~\cite{giamarchi_general_ref}~\footnote{Thus, 1D dipolar
  quantum gases are amenable to a variety of interesting effects, as
  {\it e.g.} spin-charge separation (see A. Kleine and C. Kollath and
  I. McCulloch and T. Giamarchi and U. Schollwoeck, arXiv:0706.0709 (2007))}. 
We have more recently predicted
that dipolar bosonic quantum gases confined in quasi-1D geometries can reach
correlation regimes well beyond those of the (already strongly
correlated) Tonks-Girardeau (TG)
gas~\cite{TG}, crossing over to a Dipolar-Density-Wave (DDW)
state at very large densities $n$ on the scale of the potential 
range, where the atoms arrange into an ordered state regularly
spaced by $n^{-1}$~\footnote{By dipolar-density-wave we mean a quasi-ordered state
very much analogous to a charge-density-wave}. 
By a back-to-back comparison with Reptation Quantum Monte Carlo (RQMC)~\cite{RQMC}
simulational data, we have shown that at the level of the static
structure factor the crossover can be described by a Luttinger-liquid
theory with exponent $K<1$ continuously decreasing from $K=1$ $nr_0\to 0$ to
$K\to 0$ as $nr_0\to\infty$. Finally, we have predicted the corresponding signatures     
in the collective excitations of the trapped gas~\cite{pedri_sub}.  

Beyond the evidence emerging from the static structure of the fluid, a clear-cut
demonstration of Luttinger behavior requires further understanding
of the excitations in the homogeneous dipolar gas.
In particular, answers to two relevant questions are not obvious from
the beginning. First, whether roton-like
excitations may show up in the dipolar gas at finite wavevectors. 
Second, whether the quantum
fluctuations of the phonon field prevent the existence of long-range
order at large densities, namely whether 
the crystal order parameter vanishes in the thermodynamic limit.
In fact, exploiting the uncertainty-principle instead of the
Bogolubov inequality, Pitaevskii and Stringari~\cite{pita_stringa_QF}
have worked out an extension of the Hohenberg-Mermin-Wagner theorem~\cite{HMW,Martin}
which yields more accurate upper bounds to the size of the order
parameter at zero temperature, where the quantum fluctations
dominate. When applied to specific systems, the inequality may allow 
to rule out the existence of
long-range order, as in the case of, {\it e.g.}, 1D antiferromagnets and
crystals~\cite{pita_stringa_QF}. Both questions above would have a
definite answer if the system
were in a Luttinger-liquid state, for which there is no long-range order nor
roton minimum. 
 
In this Brief Report, we find that this is indeed the case, after
computing by RQMC the low-energy excitation spectrum up to eight reciprocal lattice
vectors $G_m/n=2\pi m$ 
in the whole crossover. The evolution of quasi-long-range order 
from the TG to the DDW state emerges as a progressive closing
of the gaps in the excitation spectrum with increasing 
the order $m$. By the same token, we  demonstrate the absence 
of a roton minimum at
$2\pi$ in the whole crossover and that dynamical effects play a significant 
role 
in building the Luttinger state. Our results, analyzed by means of the
uncertainty-principle inequality~\cite{pita_stringa_QF}, also 
rule out the existence of long-range
order in this super-strongly correlated quantum gas and confirm that
the 1D dipolar gas is in a Luttinger-liquid state.  

{\it The model and the RQMC method.-}
We model the 1D dipolar Bose gas by considering $N$ 
atoms with mass $M$ and permanent dipoles moments
arranged along and orthogonal to a line, yielding
purely repulsive interactions. The Hamiltonian is
\begin{equation}
H=-\frac{1}{r_s^2}
\sum_{i} \frac{\partial^2}{\partial x_i^2}+
\frac{1}{r_s^3} \sum_{i<j}\frac{1}{|x_i-x_j|^3} \; .
\label{eq:HdipoleRy}
\end{equation}
in effective Rydberg units 
$Ry^*={\hbar^2}/({2 M r_0^2})$. The effective Bohr radius $r_0\equiv M C_{dd}/(2
\pi \hbar^2)$ is expressed in terms of the interaction strength
$C_{dd}=\mu_0\mu_d^2$ for magnetic and 
$C_{dd}=d^2/\epsilon_0$ for electric dipoles~\footnote{$\mu_d$ and $d$ are the magnetic and electric
dipole moments and $\mu_0$ and $\epsilon_0$ are the vacuum permittivities}.
The dimensionless parameter $r_s=1/(n  r_0)$ determines the interacting
regime in terms of $r_0$ and of the linear density $n$. Since the potential-to-kinetic energy ratio scales
as $1/r_s=nr_0$, large
densities yield to strong correlations, at variance with Coulomb systems.   

We determine the ground-state properties and the excitation spectrum
by resorting to the Reptation Quantum Monte Carlo
technique~\cite{RQMC}. This is in essence a path-integral method at zero
temperature, where the ground-state distribution is directly sampled
in the internal part of the path. Thus, the computation of the
structure of the fluid and of the 
imaginary-time correlation functions for suitable long projection times is
conceptually straightforward and practically easy, since possible
biases arising from mixed averages are ruled out by definition.   
In particular, from an analysis of the imaginary-time density-density correlation
function we determine the low-energy excitation spectrum while the parameter
$nr_0$ spans the whole crossover from the TG to the DDW state. 

We use a trial wave-function that is a product of two-body Jastrow factors 
$\psi_{trial}=\prod_{i<j} e^{u(|z_i-z_j|)}$. As we are interested in 
long-range behavior, we actually take the Luttinger-liquid expression
\begin{equation}
\psi_{trial}({R})
\propto \prod_{i<j} |\sin \frac \pi L (x_i - x_j) |^{1/K}\; .
\end{equation}
which in the low-density limit implies
$K=1$~\cite{dipolar_citro_pra_rc} and recovers the wave-function of spinless non-interacting fermions. 
Different choices of the wave functions, such as the product of gaussians centered 
on the lattice sites $R_m=mn^{-1}$, result into different time-step
extrapolations, but eventually lead to negligible differences in
the computation of the static and dynamic structure factors.

A few technical details are in order. 
We perfom simulations for different values of the number $N$ of bosons
in a square box with periodic boundary conditions, namely
$N=40,60,80,100$, reaching in selected cases $N=200$. We check that we
are able to take care of finite-size 
effects by summing the interactions over ten simulation
boxes. Finally, the energies are extrapolated to their thermodynamic limit
after removing the time-step dependence. 

The resulting energy per particle $\varepsilon(nr_0)/Ry^*$ as a
function of $nr_0$ has been provided in ~\cite{pedri_long_sub,pedri_sub},  
together with an accurate analytical form of it, useful for further
applications. We remark here that $\varepsilon(nr_0)/Ry^*$ 
recovers the known limiting behaviors $\varepsilon(nr_0)/Ry^*\sim
(\pi^2/3)(nr_0)^2$ for $nr_0 \ll 1$ in the TG regime and
$\varepsilon(nr_0)/Ry^* \sim \zeta(3)(nr_0)^3$ for $n r_0 \gg 1$ in
the DDW limit~\cite{dipolar_citro_pra_rc}.  
We set from now on units of $n^{-1}$ for lengths and of $Ry^*$ for
energies. 

{\it Low energy excitations from dynamical structure factor.-} 
We determine the low-energy excitations after computing the
imaginary-time correlation function of the density operator~\cite{stefania_2dbose} 
$\rho_q=\sum_i exp(-i  \vec{q}\cdot \vec{r_i(\tau)})$ creating a
density fluctuation with wave vector $q$, that is
$F(q,\tau)=\langle \rho_q(\tau)\rho_q^{\dagger}(0) \rangle/N$,
where the sum in $\rho_q$ spans over the number of particles $N$
located at position $\vec{r_i}$ and $\tau$ is the imaginary time. 
$F(q,\tau)$ is related to the dynamical structure factor by
$F(q,\tau)=\int^{\infty}_0 d\omega \exp(-\omega\tau) S(q,\omega)$,
yielding the static structure factor for $\tau=0$, namely
$S(q)=F(q,0)=\int^{\infty}_0 d\omega S(q,\omega)$. 

\begin{figure}[tb]
\includegraphics[width=85mm]{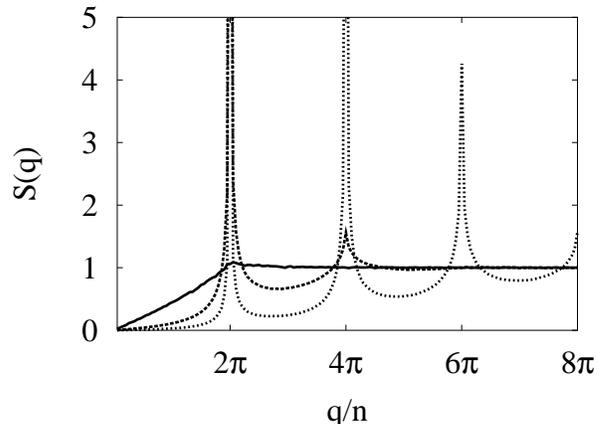}
\vspace{-5mm} \caption{$S(q)$ in dimensionless units
for a dipolar gas with $N=40$
particles and different values of $n r_0=0.01,\ 50$ and $1000$. 
Decreasing slopes as $q\to 0$ and the emergence of additional
peaks correspond increasing $nr_0$ values.} \label{fig:sofq}
\end{figure}

To give a system overview, we first display the $S(q)$ in Fig.~\ref{fig:sofq} as previously
determined~\cite{dipolar_citro_pra_rc} by RQMC for various densities between the
TG and DDW limits. Peaks 
at $q/n=2\pi m$ ($m$ integer) progressively disappear with decreasing density as 
the dipolar gas approaches the spinless fermionic liquid.

We track this smooth evolution by investigating the low-energy
excitations as extracted from the dynamical structure factor
$S(q,\omega)$. From the expression
\begin{equation}
S(q,\omega)=\sum_n|\langle n|\rho_q|0 \rangle|^2
\delta(\omega-\omega_n)\; ,
\label{excitations}
\end{equation}
we estimate the energy dispersion of the collective excitations by 
fitting the imaginary-dependence of $F(q,\tau)$ as a sum of exponentials 
$\sum_i A_i(q) e^{\omega_i(q)\tau}$ corresponding to 
multiple modes. Then, the fit yielding the
best $\chi^2$ value is chosen.    

\begin{figure}[tb]
\includegraphics[width=85mm,angle=-90]{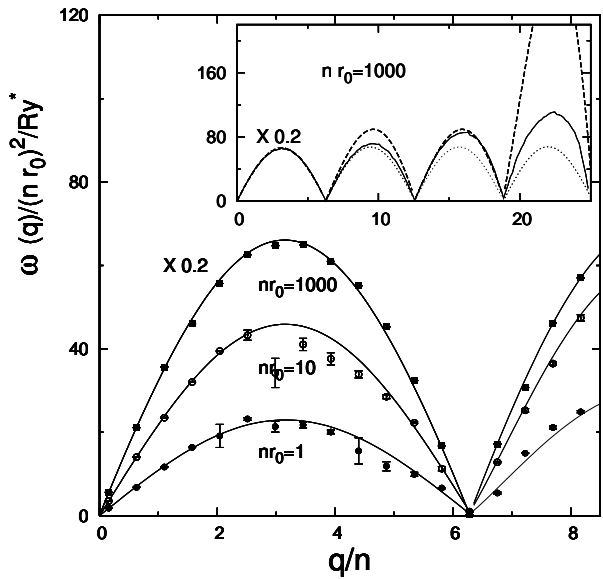}
\vspace{-1mm} \caption{Lowest excitation energies $\omega(q)$ in
$Ry^*$ units and scaled by $(nr_0)^2$, 
for a dipolar gas with $N=40$ and different values of $nr_0=1, 10$ and $1000$ 
as in the legend. Symbols with error bars
represent the RQMC data extracted from (\ref{excitations}), 
the solid line is a guide to the eye. 
The curve at $nr_0=1000$ is depressed by 
a factor of 5 for graphical reasons. Inset: zoom on the 
$\omega(q)$ at $nr_0=1000$ up to $q/n=8\pi$ for different $F(q,\tau)$ 
models: multimode (solid) and Feynman (dashed) approximation. 
Dotted line: periodic replica of the first bump.} 
\label{fig:omegak}
\end{figure}

Fig.~\ref{fig:omegak} displays the resulting RQMC energy dispersion
$\omega(q)$, for the case with 
$N=40$ at $nr_0=1,10$ and $1000$ namely in the 
low, intermediate and very high density regimes. In spite of the
finite size effects, the overall qualitative behavior is already
clear. The phonon softens at low $q$-values, 
while the density decreases. On the other hand, 
the gap at $2\pi$ seems to be always closed at all densities.
As represented in the inset by the dotted curve, 
even at $nr_0=1000$ the RQMC excitation spectrum (solid line) 
is very different from what could be obtained 
by replicating the portion from $q=0$ to $q=2\pi$ (dotted line), 
indicating a non periodic structure. The dashed line instead,
represents $\omega(q)$ obtained
from the Feynman relation $\omega(q)=\epsilon(q)/S(q)$, which 
provides only an upper bound to $\omega(q)$ in terms of
the static structure factor $S(q)$ and of the kinetic energy
$\epsilon(q)=\hbar^2q^2/2m$. Indeed, the Feynman relation is clearly
seen to overestimate the RQMC value for $q>2\pi$, while it seems to
account for the $\omega(q)$ between $q=0$ and $q=2\pi$.    
Considering that the Feynman relation is expected to yield better
results as $q/n\to 0$, we can also anticipate
that at lower densities the $q$ range where the Feynman relation is
reliable will shrink (see below).  

\begin{figure}[tb]
\includegraphics[width=75mm,angle=-90]{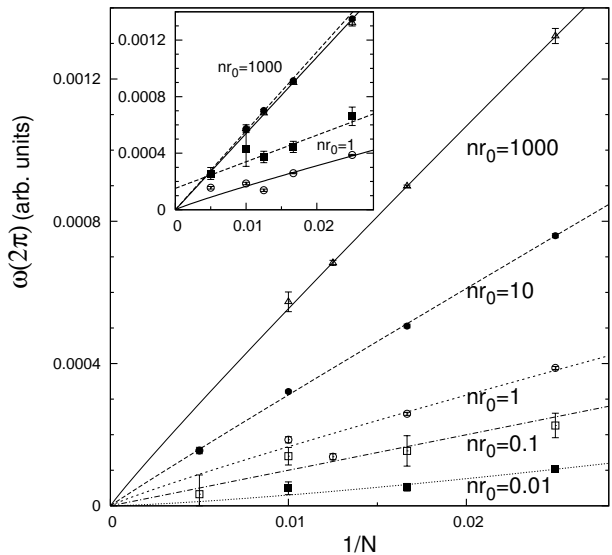}
\vspace{-1mm} \caption{$1/N$ scaling of $\omega(q=2\pi)$ in arbitrary units
at $n r_0=0.01,\ 0.1,\ 1,\ 10$ and $1000$
as in the legend. Symbols with error bars:
RQMC data. Solid lines: fit to the data. Inset: zoom of the RQMC data
and related fits for the cases
with  $nr_0=1$ and $1000$. Filled symbols and dashed lines: Feynman relation
(single-mode approximation). Open symbols and solid lines: two-mode
analysis yielding the best $\chi^2$.} 
\label{fig:omega_2pi_L}
\end{figure}

A quantitatively reliable measure of the gap sizes requires an accurate size
effect analysis. Fig.~\ref{fig:omega_2pi_L} displays the $1/N$ scaling 
of $\omega(q/n=2\pi)$ for $nr_0=0.01, 0.1, 1, 10$ and
$1000$. The fit to the RQMC data (symbols in the figure) 
yields the linear scaling $\omega_N(q=2 \pi)=c(nr_0)/N$
with the constant $c(nr_0)$ being an increasing function of $nr_0$.  
Thus, $\omega(q/n=2\pi)\to 0$ as $1/N\to 0$, the gap is 
clearly closed at all densities, demonstrating the absence of a roton
minimum. This is not the case if we use the
Feynman relation, corresponding to a single-mode approximation in 
Eq.~(\ref{excitations})~\cite{RQMC}. At the intermediate-to-low
density $nr_0=1$ the fit yields a finite gap value (see the inset). Since the Feynman
relation is built up from static quantities, we may conclude that
dynamical effects, as embodied in the multimode analysis of
(\ref{excitations}) play a significant role, leading to
qualitatively different conclusions. 
A similar multimode analysis performed at $q=2\pi m$, $m>2$ shows the existence of open gaps, which progressively close while the
quasi-ordered state is approached. 

{\it Absence of long-range order.-}
Using these results, we can derive a strict upper bound for 
the order parameter of the solid 
$\rho_{\vec{q}}=N^{-1}\langle\sum_m\exp(i\vec{G}\cdot\vec{r}_m)\rangle$ with $\vec{G}$ a vector of
the reciprocal lattice, and rigorously
test the qualitative conclusions from the inset of
Fig.~\ref{fig:omegak}, namely that no long-range order may exist  
in our 1D dipolar quantum gas. We closely follow 
the derivation of Pitaevskii and
Stringari~\cite{pita_stringa_QF}. By applying the
uncertainty-principle inequality
$\langle{A^\dagger,A}\rangle\langle{B^\dagger,B}\rangle\geq|\langle[A^\dagger,B]\rangle|^2$ 
to the operators $A=\hat{\rho}_{\vec{q}+\vec{G}}$ and
$B=\partial\hat{\rho}_{\vec{q}}/\partial t$ one has 
$S(\vec{q}+\vec{G})\int d\omega
\omega^2S(\vec{q},\omega)\geq\frac{1}{4m^2}\rho^2_{\vec{G}}(\vec{q}\cdot(\vec{q}+\vec{G})^2))$. 
From the RQMC data we know that as $q\rightarrow 0$, $S(\vec{q}+\vec{G})\rightarrow
|q|^{2K-1}$ , while the second moment of $S(q,\omega)$ vanishes as
$|q|^3$. Thus, the order parameter in the long wavelength limit vanishes as $\rho^2_{\vec{G}}\leq
q^{min(2K,1)}$ with $K\geq 0$. Thus no long-ranger order may exist unless
$K=0$, which is however the limit of infinite density.

{\it Dynamical response function and Luttinger-liquid analysis.-}
These results can be analyzed within the Luttinger-liquid theory. 
We want to calculate the imaginary-time $\tau$ correlation function
$\tilde{F}(x,\tau)=\langle T_\tau e^{i 2\phi (x,\tau) } e^{-2i \phi(0,0)}
\rangle$, on a finite size system of length $L$. It is
known~\cite{giamarchi_book_1d} from bosonization that
$\tilde{F}(x,\tau)= {(\pi \alpha/L)^{2K}}/{\left[\sinh^2\left(\frac{\pi u \tau}{L}\right) + \sin^2\left(\frac{\pi x} L \right)\right]^K}$
valid in the long-time (low-energy) limit $u\tau\gg\alpha$ where the  
$\alpha$ is a short-distance cutoff of the order of $n^{-1}$, $u$ is the velocity of the excitations and $K$ the Luttinger exponent.  
After Fourier transforming $\tilde{F}(x,\tau)$  in $q$-space with 
$q=2\pi j/L$, we get:
\begin{eqnarray}
  \label{eq:transformed-corr}
&&  F(q,\tau)= \left(\frac{\pi \alpha} L\right)^{2K} L 2^{2K+1-j}
  \frac{ e^{-\frac{2\pi u \tau}{L} (K+j)}} {\left(1+e^{-\frac{2\pi u
          \tau}{L}}\right)^{K+j}} \times\\
&&[{\Gamma(j+K)}][{\Gamma(K)
  \Gamma(j+1)}]^{-1}{}_2F_1\left(K+j,K;j+1;e^{-\frac{2\pi u
      \tau}{L}}\right)\nonumber\; ,
\end{eqnarray}
where ${}_2F_1$ and $\Gamma$ are the Hypergeometric and Euler functions. 
 
Using an Ansatz $e^{-\tau \omega_1(q)}$ to fit the long-time behavior 
$F(q,\tau\to\infty)\sim e^{-\frac{2\pi u \tau}{L}  (K+j)}$ we get $
  \omega(q)= {2\pi u K}/{L} + u |q|$,
where we have used the even parity of the response function.
Thus, there should be no roton gap at $q=2\pi$ in the infinite size
limit. For finite size, an apparent roton gap (vanishing as $1/L$) can be
seen. This gap can be traced to the zero mode contribution to the
correlation functions. All the fits to the RQMC data presented in
Figs.~\ref{fig:omegak}-\ref{fig:omega_2pi_L} reproduce remarkably well
this $1/L$ scaling~\footnote{At fixed density, $1/N$ and $1/L$
  scaling are equivalent.}, and are consistent with our previous
findings on the density dependence of Luttinger-$K$ exponent $K(n)$~\cite{dipolar_citro_pra_rc}.\\ 
{\it Conclusions.-} In conclusion, the analysis of the RQMC
simulational data neatly lead to two main conclusions, namely that 
there are no roton excitations 
appearing at the first star of the reciprocal lattice and that no long
range order may exist in the whole
crossover from the TG gas at low density to the
quasi-ordered DDW state at high densities. The RQMC data analysis 
is in remarkable agreement with what expected for a super-strongly
correlated Luttinger-liquid state, with the inclusion of significant
dynamical effects. The realization of 1D dipolar quantum (molecular) gases in the TG to
the DDW regime is within reach of current experimental
efforts~\cite{dipolar_citro_pra_rc} and thus our predictions on the
excitation spectrum and on the absence of
the roton minimum, can be tested in future experiments by means of {\it
  e.g.} Bragg spectroscopy techniques~\cite{stamper-kurn99_bragg_bec}.  

{\it Acknowledgments}. We are especially grateful to the Scuola
Normale Superiore to have provided ideal conditions for the
realization of large parts of this work, and to G. La Rocca for interesting 
discussions and support. We also thank S. Stringari and 
L. Pitaevskii to have pointed out the use of the uncertainty-principle 
inequality.  


\end{document}